\documentclass[manuscript]{aastex61}
\usepackage{amsmath}
\usepackage{color}

\newcommand{\ppar}{p_{\parallel}}
\newcommand{\pper}{p_{\perp}}
\newcommand{\cc}{\textrm{cm}^{-3}}
\newcommand{\AU}{\textrm{AU}}

\begin{document}

\title{Plasma Energization in Colliding Magnetic Flux Ropes}

\author{Senbei Du}
\affiliation{Department of Space Science, University of Alabama in Huntsville, Huntsville, AL 35899, USA}
\affiliation{New Mexico Consortium, Los Alamos, NM 87544, USA}

\author{Fan Guo}
\affiliation{Los Alamos National Laboratory, Los Alamos, NM 87545, USA}
\affiliation{New Mexico Consortium, Los Alamos, NM 87544, USA}

\author{Gary P. Zank}
\affiliation{Department of Space Science, University of Alabama in Huntsville, Huntsville, AL 35899, USA}
\affiliation{Center for Space Plasma and Aeronomic Research (CSPAR), University of Alabama in Huntsville, Huntsville, AL 35805, USA}

\author{Xiaocan Li}
\affiliation{Los Alamos National Laboratory, Los Alamos, NM 87545, USA}

\author{Adam Stanier}
\affiliation{Los Alamos National Laboratory, Los Alamos, NM 87545, USA}

\begin{abstract}

Magnetic flux ropes are commonly observed throughout the heliosphere, and recent studies suggest that interacting flux ropes are associated with some energetic particle events. In this work, we carry out 2D particle-in-cell (PIC) simulations to study the coalescence of two magnetic flux ropes (or magnetic islands), and the subsequent plasma energization processes. The simulations are initialized with two magnetic islands embedded in a reconnecting current sheet. The two islands collide and eventually merge into a single island. Particles are accelerated during this process as the magnetic energy is released and converted to the plasma energy, including bulk kinetic energy increase by the ideal electric field, and thermal energy increase by the fluid compression and the non-ideal electric field. We find that contributions from these different energization mechanisms are all important and comparable with each other. Fluid shear and a non-gyrotropic pressure tensor also contribute to the energy conversion process. For simulations with different box sizes ranging from $L_x \sim 25$--$100 d_i$ and ion-to-electron mass ratios $m_i / m_e = 25$, 100 and 400, we find that the general evolution is qualitatively the same for all runs, and the energization depends only weakly on either the system size or the mass ratio. The results may help us understand plasma energization in solar and heliospheric environments.

\end{abstract}


\section{Introduction} \label{sec:intro}

Magnetic islands are 2D magnetic structures, characterized by closed-loop-like field lines. In three dimensional space, when an out-of-plane magnetic field is included, the field lines form helical structures called magnetic flux ropes. Magnetic islands and flux ropes are frequently observed by spacecrafts, such as in the Earth's magnetotail \citep{Wang2016Nat}, in the solar atmosphere \citep{Takasao2016ApJ} and near the heliospheric current sheet (HCS)\citep{Cartwright2010JGR}. They may play an important role in particle acceleration and energization of solar and heliospheric plasmas \citep[e.g.,][]{Khabarova2017ApJ}.

The formation and evolution of magnetic flux ropes are closely related to magnetic reconnection, which is a sudden change in the magnetic field line configuration, efficiently converting magnetic free energy into plasma energy. Previous simulations suggest that magnetic islands are formed by multiple reconnection along a current sheet, or in a 2D turbulent reconnecting flow \citep[e.g.,][]{Servidio2009PRL}. In the solar wind, magnetic flux ropes are found to be located near the HCS \citep{Cartwright2010JGR, Khabarova2015ApJ}, where magnetic reconnection is likely to occur, and they may be a result of the cascade of quasi-2D MHD turbulence \citep{Greco2009ApJL, Zank2017ApJ}. Indeed, the current paradigm for solar wind turbulence is that it is comprised of a majority of 2D component superimposed with a minority slab component \citep{Zank1992JGR, Zank1993PFA, Bieber1996JGR, Zank2017ApJ}, at least in those regions for which the plasma beta $\sim 1$ or $< 1$. \citet{Zheng2018ApJL} recently developed a database of small-scale magnetic flux ropes, and their results appear to be consistent with this view, though they do not discuss in particular the coincidence with the HCS.

A very interesting aspect of magnetic flux ropes is their potential to accelerate charged particles. Spacecraft observations find that direct magnetic reconnection is not an efficient particle accelerator in the solar wind \citep{Gosling2005GRL}. On the other hand, recent observations indicate some energetic particle events are associated with the crossing of the HCS and small-scale magnetic flux ropes \citep{Khabarova2015ApJ, Khabarova2016ApJ, Khabarova2017ApJ}. In particular, \citet{Khabarova2017ApJ} reexamined the ``Gosling event'' \citep{Gosling2005GRL} and 126 related events over a larger time and more extended spatial range and energies. They found evidence of an energetic particle population, most likely accelerated by turbulence/magnetic island-related structures generated by the initial reconnection event. In these regions, magnetic flux ropes undergo dynamic interactions through magnetic reconnection. Particles that are trapped in such regions can be accelerated more efficiently than in an isolated reconnection exhaust. Numerical simulations have suggested several mechanisms for particle acceleration \citep[e.g.,][]{Drake2006Natur, Drake2006GRL, Drake2013ApJL, Oka2010ApJ, Le2012PoP}. The basic processes include first order Fermi acceleration due to magnetic island contraction, and direct acceleration by the reconnection electric field generated during the merging of two adjacent magnetic islands. \citet{Zank2014ApJ_rec} developed a theoretical particle transport equation that describes the particle acceleration in a ``sea'' of interacting magnetic islands, incorporating the abovementioned basic acceleration mechanisms. \citet{leRoux2015ApJ} develop a more sophisticated transport equation from a quasi-linear theory (QLT) approach, where the energization is associated with guiding center drift motions. With the exception of one term associated with the variance of magnetic-island-induced electric field, there is a one-to-one correspondence between the more physically-based derivation \citep{Zank2014ApJ_rec} and the QLT derivation \citep{leRoux2015ApJ}. The energization mechanisms for the two equations discuss exactly the same effects but in terms of ``field line contraction'' versus ``guiding center motions''.
The \citet{Zank2014ApJ_rec} theory is recently applied to explain a \emph{Ulysses} observation of an unusual energetic particle flux enhancement near 5 AU that can be associated with small-scale magnetic island dynamics behind an interplanetary shock \citep[][in press]{Zhao2018}. The theoretical solution agrees quantitatively with the observed energetic proton flux amplification and spectral evolution.
Numerical simulations find that the curvature drift and the parallel electric field have the most important contribution to the energization during magnetic reconnection \citep{Dahlin2014PoP, Guo2014PRL, Guo2015ApJ, Li2015ApJL, Li2017ApJ}. Compressibility is an important factor in plasma energization, as it is related to both the magnetic island contraction mechanism and drift motions, but its role is still under debate. For example, some previous simulations suggest that island coalescence is an incompressible process \citep{Fermo2010PoP, Drake2010ApJ}. \citet{Drake2013ApJL} derive a particle transport equation based on the assumption of incompressible island coalescence that increases parallel particle energy and reduces perpendicular energy. On the contrary, both parallel and perpendicular energy can increase in a compressible flux rope, leading to a first-order Fermi acceleration. Both processes (compressible and incompressible) are included in the transport equations derived by \citet{Zank2014ApJ_rec} and \citet{leRoux2015ApJ}. Recent simulations \citep{Li2018ApJ} and theoretical work \citep[][in press]{leRoux2018} find that the fluid compression is the dominant energization mechanism, especially when a guide field is absent or weak. As discussed above, the energization of particles by magnetic flux ropes might be viewed as an underlying mechanism for the acceleration of ions and electrons by low frequency MHD turbulence corresponding to that found in the solar wind.

Due to computational restrictions, kinetic simulations can only be applied to relatively small systems with unrealistic mass ratios and Alfv\'{e}n speeds. In a space physics and astrophysics context, the measurable scale sizes can be much larger than the scales considered in kinetic simulations. For example, small magnetic flux ropes observed in the solar wind have a typical scale size of $\sim 0.001$--$0.01 \AU$ \citep{Cartwright2010JGR, Zheng2018ApJL} at a heliocentric distance of $1 \AU$ (smaller magnetic flux ropes may exist but are not observed due to limitations of the data cadance). On assuming a proton number density of $n_i = 10 \cc$, the size of small flux ropes corresponds to $\sim 2000$--$20000 d_i$, where $d_i = c / \omega_{pi}$ is the ion inertial length. A system of interacting magnetic islands with such sizes is beyond the capability of most current kinetic simulations. On the other hand, magnetic flux ropes in the Earth's magnetosphere have been observed down to the scale of a few ion inertial lengths owing to the availability of high-sampling-rate instruments \citep[e.g.,][]{Wang2016Nat, Huang2016JGR}. Due to the ubiquitous presence of magnetic flux ropes at different scales, it is useful to study how kinetic simulation results scale to larger and more realistic applications in the heliosphere.

In this study, we perform 2D fully kinetic particle-in-cell (PIC) simulations of the coalescence of two magnetic flux ropes. We analyze the energy conversion from a fluid perspective, and show the contribution of three important energy conversion terms: the ideal electric field, the non-ideal electric field, and the fluid compression. We explore a range of system sizes and show that our results depend only weakly on the size.

\section{Simulations} \label{sec:sim}

The simulations use the VPIC code \citep{Bowers2008PoP}, which solves Maxwell's equations for electromagnetic fields and the relativistic equation of motion for particles. Four simulation runs were carried out to explore different simulation parameters. The initial configuration consists of two magnetic islands embedded in a reconnecting current sheet, illustrated in the top panels of Figure \ref{fig:example}. A similar setup has been used in several previous simulation studies \citep[e.g.,][]{Stanier2015PRL, Stanier2015PoP, Stanier2017PoP}. The magnetic field is given by
\[ B_x = \frac{B_0 \sinh(z/L)}{\cosh(z/L) + \varepsilon\cos(x/L)},\, B_z = \frac{\varepsilon B_0 \sin(x/L)}{\cosh(z/L) + \varepsilon\cos(x/L)},\, B_y = \frac{B_0\sqrt{1-\varepsilon^2}}{\cosh(z/L) + \varepsilon\cos(x/L)}, \]
where $L$ is the half thickness of the current sheet, which also determines the system size, and $\varepsilon$ is a measure of the island size. This setup ensures that the initial condition is force free, i.e., $\boldsymbol{J\times B} = c\boldsymbol{(\nabla \times B) \times B}/(4\pi) = 0$. The simulation box is set to $x \in [0, 4\pi L]$, $z \in [-\pi L, \pi L]$, and $L$ ranges from $2d_i$ to $8d_i$, $d_i = c / \omega_{pi}$ is the ion inertial length. We set $\varepsilon = 0.4$ in all simulation runs. A periodic boundary condition is applied to the $x$ direction, and in the $z$ direction conducting field\slash reflective particle boundaries are used. We consider only an electron-proton plasma, and the mass ratio $m_i/m_e$ is set to 25 for the first three cases. For the smallest domain size, we also test higher mass ratios 100 and 400. The initial electron and proton temperatures are uniform throughout the simulation domain with $kT_e = kT_i = 3.75\times 10^{-3} m_e c^2$. The magnetic field strength $B_0$ is determined by the $\omega_{pe}/\Omega_{ce}$ value, where $\omega_{pe} = \sqrt{4\pi n_0 e^2 / m_e}$ and $\Omega_{ce} = m_e c / eB_0$ are the electron plasma frequency and electron gyrofrequency, respectively. We set $\omega_{pe}/\Omega_{ce} = 2$ in our simulations. The particle number density profile is also uniform initially. The initial plasma beta is then given by $\beta_e = \beta_i = n_0 kT_e / (B_0^2/2) = 0.03$. We make five simulation runs to test different system sizes and mass ratios. About 400 macro-particles per cell are used for each species for all but the 400-mass-ratio run, where 1600 macro-particles per cell are used to suppress the numerical heating (discussed in Section \ref{sec:general}). The varying parameters are listed in Table \ref{tab:par}. Note that in our simulations, all the characteristic velocities are small relative to the speed of light. Therefore, it is appropriate to use non-relativistic equations in our analysis.

\begin{deluxetable}{ccCCc}[htb!]
\tablecaption{Simulation parameters.}
\tablehead{
\colhead{Run \#} &
\colhead{$m_i/m_e$} &
\colhead{$L_x \times L_z(d_i)$} &
\colhead{$N_x \times N_z$} &
\colhead{$t_{end}(\Omega_{ci}^{-1})$}
}
\startdata
1  & 25  & 8\pi \times 4\pi    & 1024 \times 512  & 200 \\
2  & 25  & 16\pi \times 8\pi   & 2048 \times 1024 & 400 \\
3  & 25  & 32\pi \times 16\pi  & 4096 \times 2048 & 800 \\
4  & 100 & 8\pi \times 4\pi    & 2048 \times 1024 & 200 \\
5  & 400 & 8\pi \times 4\pi    & 4096 \times 2048 & 200 \\
\enddata
\tablecomments{$L_x \times L_z$ is the simulation domain size, $N_x \times N_z$ is the grid resolution, $t_{end}$ is the total simulation time.} \label{tab:par}
\end{deluxetable}

\section{Results} \label{sec:result}
\subsection{General Evolution} \label{sec:general}
\begin{figure}[!htb]
\centering
\plotone{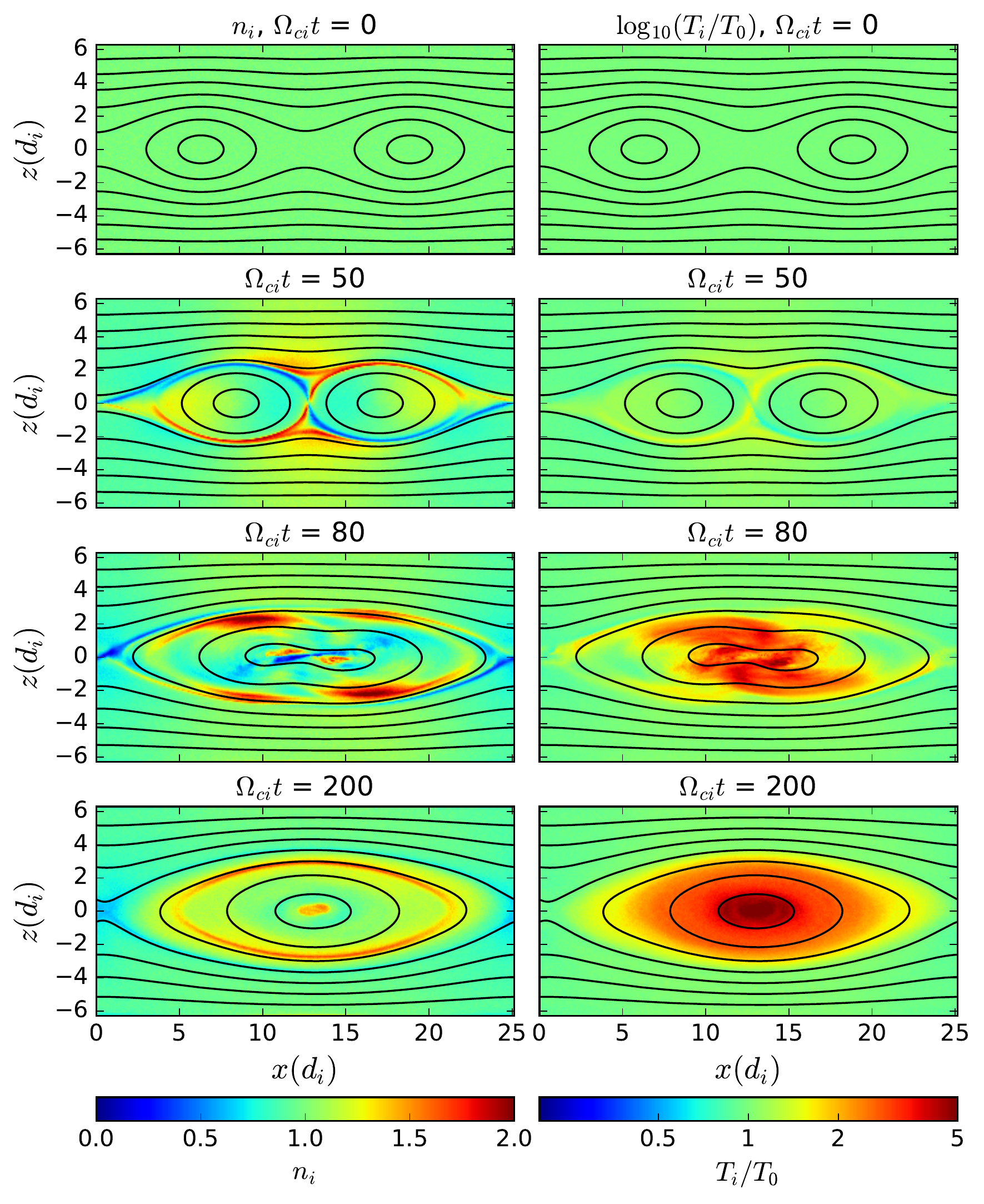}
\caption{Snapshots from Run 4 at four instances. Left panels show the ion number density, and right panels show the ion temperature profile (plotted logarithmically). In-plane magnetic field lines are superimposed.} \label{fig:example}
\end{figure}

The simulation domain consists initially of two magnetic islands embedded in a current sheet. Two reconnection X-lines are identified, one at the center, the other at the boundaries in the $x$-direction. As the reconnection occurs, the islands move towards each other and collide, and eventually merge into a single island. Small perturbations are introduced to break the symmetry of the initial setup and initiate the island coalescence instability. To illustrate the general evolution, we show several snapshots from Run 4 in Figure \ref{fig:example}, where the ion number density and temperature are color-coded and magnetic field lines are superimposed. As the two islands merge,
particles gain energy from the magnetic reconnection process indicated by the increased electron and ion temperature (the ion temperature is shown in Figure \ref{fig:example}). Note that the temperature increase is an indication of plasma energization, but it is not very informative about the accelerated component of the particle distribution function. The newly merged island appears to oscillate before it settles to a nearly stationary elongated state. After examining all simulation cases, we find neither the system size nor the mass ratio has a qualitative impact on the general evolution of the system. Previous simulations \citep{Stanier2015PRL, Stanier2017PoP} find that a large system size will result in a smaller reconnection rate (in the high beta regime this occurred for $L \gtrsim 10 d_i$). New features such as secondary magnetic island formation and islands ``bouncing off'' each other may also arise in larger-size simulations. Due to the limited system size ($L \le 8d_i$) in this study, we cannot fully test those effects.

As the two magnetic flux ropes merge, the magnetic energy is released and the particle kinetic energy increases as a result. The energy budget is listed in Table \ref{tab:energy}.
The presence of numerical heating is a known issue in PIC simulations, where the total energy of a system increases with time. The VPIC code implements an energy conserving algorithm that improves the total energy conservation efficiently \citep{Bowers2008PoP}. However, in this situation the true energy conservation can only be achieved when the time step approaches zero.
We examine this artificial effect by tracing the total energy evolution of the system. Table \ref{tab:energy} shows that the total energy increase is very small compared to energy conversion due to physical processes for all runs.
To further test the effect of numerical heating, we make 3 test runs denoted as 1a, 1b and 1c. In Run 1a, the grid resolution is doubled ($n_x \times n_z = 2048 \times 1024$); in Run 1b, the grid resolution is kept the same with Run 1 but the number of particles per cell is increased to 1600; in Run 1c, we double the grid resolution and increase the number of particles per cell to 1600 at the same time. As shown by Table \ref{tab:energy}, increasing the grid resolution or number of particles per cell reduces the numerical heating and also the released magnetic energy $\Delta \mathcal{E}_B$. The effect is more pronounced for electrons than ions. We will discuss this issue in our analysis in later sections.

The amount of released magnetic energy is relatively small (less than 5\%) compared to some previous low-$\beta$ simulations of magnetic reconnection starting from an elongated current sheet (e.g., more than 20\% in \citet{Li2017ApJ}). The reason may be (a) a large portion of the magnetic energy conversion happens during the formation of two magnetic islands in a reconnecting current sheet, which is not included in this study; (b) our initial configuration includes a relatively strong guide magnetic field component $B_y$ orthogonal to the simulation plane. Previous simulations have shown that the presence of a guide field will suppress the amount of particle energy gain \citep{Dahlin2016PoP, Li2018ApJ}. Nevertheless, particles still experience a significant energy gain in our simulations. As shown in Table \ref{tab:energy}, the total energy gain by electrons and ions ranges from $\sim 30\%$ to $80\%$.

\begin{deluxetable}{cCCCCCCCC}[htb!]
\tablecaption{Energy budget.}
\tablehead{
\colhead{Run \#} & \colhead{$\mathcal{E}_{total}^0$} & \colhead{$\Delta \mathcal{E}_{total} / \mathcal{E}_{total}^0$} &
\colhead{$\mathcal{E}_{B}^0$} & \colhead{$\Delta \mathcal{E}_{B} / \mathcal{E}_{B}^0$} &
\colhead{$\mathcal{E}_{e}^0$} & \colhead{$\Delta \mathcal{E}_{e} / \mathcal{E}_{e}^0$} &
\colhead{$\mathcal{E}_{i}^0$} & \colhead{$\Delta \mathcal{E}_{i} / \mathcal{E}_{i}^0$}
}
\startdata
1  & 134.26   & +0.03\%    & 126.50   & -4.2\%   & 4.10   & +79.9\%   & 3.65   & +57.7\% \\
1a & 134.26   & +0.01\%    & 126.50   & -4.0\%   & 4.10   & +69.9\%   & 3.65   & +58.4\% \\
1b & 134.26   & +0.01\%    & 126.50   & -3.9\%   & 4.10   & +67.2\%   & 3.65   & +59.9\% \\
1c & 134.26   & <+0.01\%   & 126.50   & -3.8\%   & 4.10   & +62.8\%   & 3.65   & +60.8\% \\
2  & 535.54   & +0.06\%    & 506.02   & -3.3\%   & 14.97  & +71.5\%   & 14.55  & +42.9\% \\
3  & 2140.66  & +0.12\%    & 2024.07  & -2.9\%   & 58.44  & +65.7\%   & 58.15  & +39.2\% \\
4  & 535.55   & +0.13\%    & 506.02   & -3.8\%   & 15.00  & +78.8\%   & 14.54  & +55.1\% \\
5  & 2140.68  & +0.12\%    & 2024.07  & -3.2\%   & 58.47  & +55.7\%   & 58.14  & +59.5\% \\
\enddata
\tablecomments{$\mathcal{E}_{total}$ is the total energy of the system, including electromagnetic energy and particle kinetic energy, $\mathcal{E}_{B}$ is the magnetic energy, $\mathcal{E}_{e}$ and $\mathcal{E}_{i}$ are the kinetic energy of electrons and ions respectively. \textcolor{magenta}{\textbf{The superscript 0 denotes the energy at the initial time.}} The unit of energy is $m_e c^2$.} \label{tab:energy}
\end{deluxetable}

\subsection{Analysis of Energy Conversion}

\begin{figure}[!htbp]
\centering
\gridline{\fig{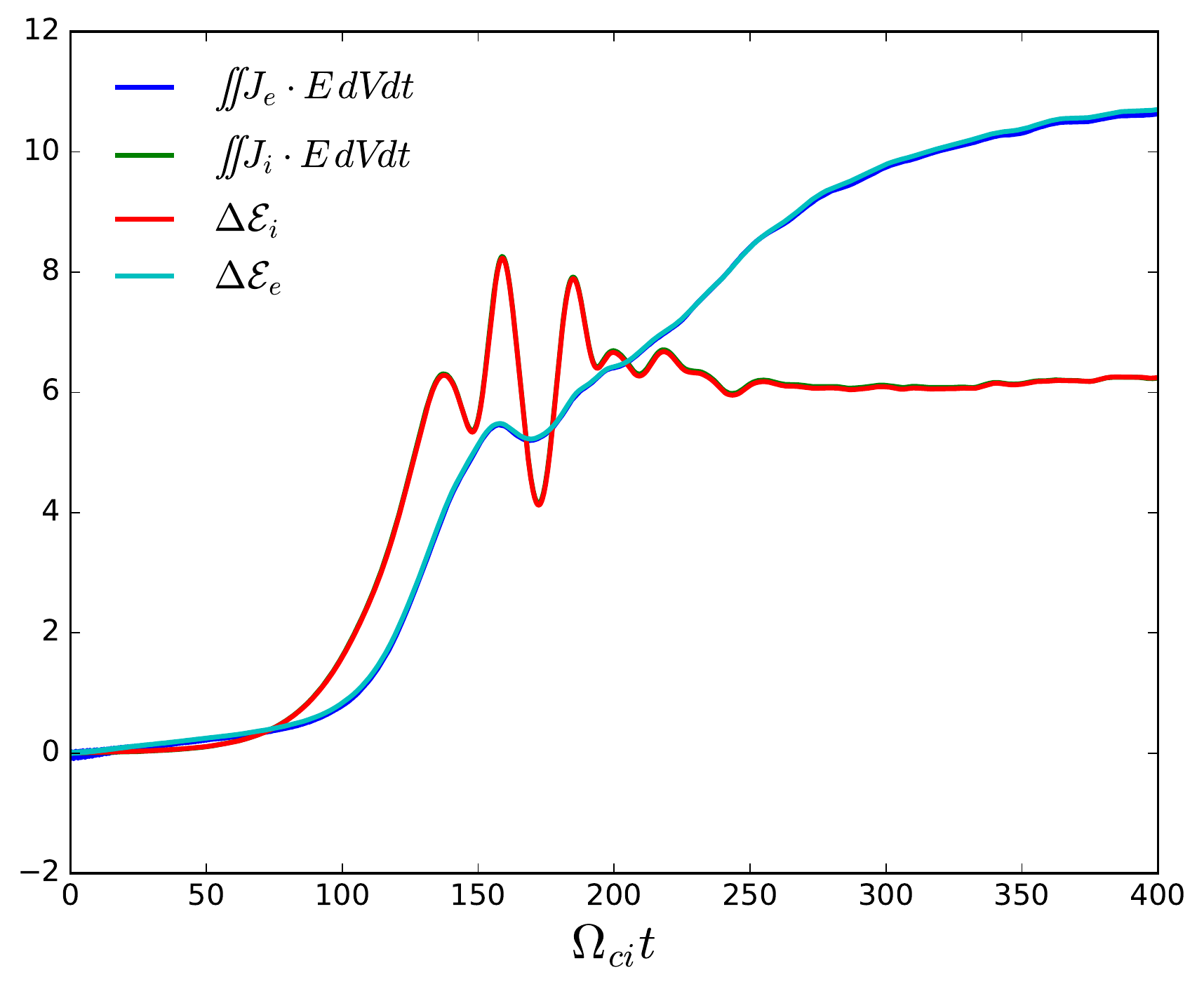}{0.5\textwidth}{(a)}%
          \fig{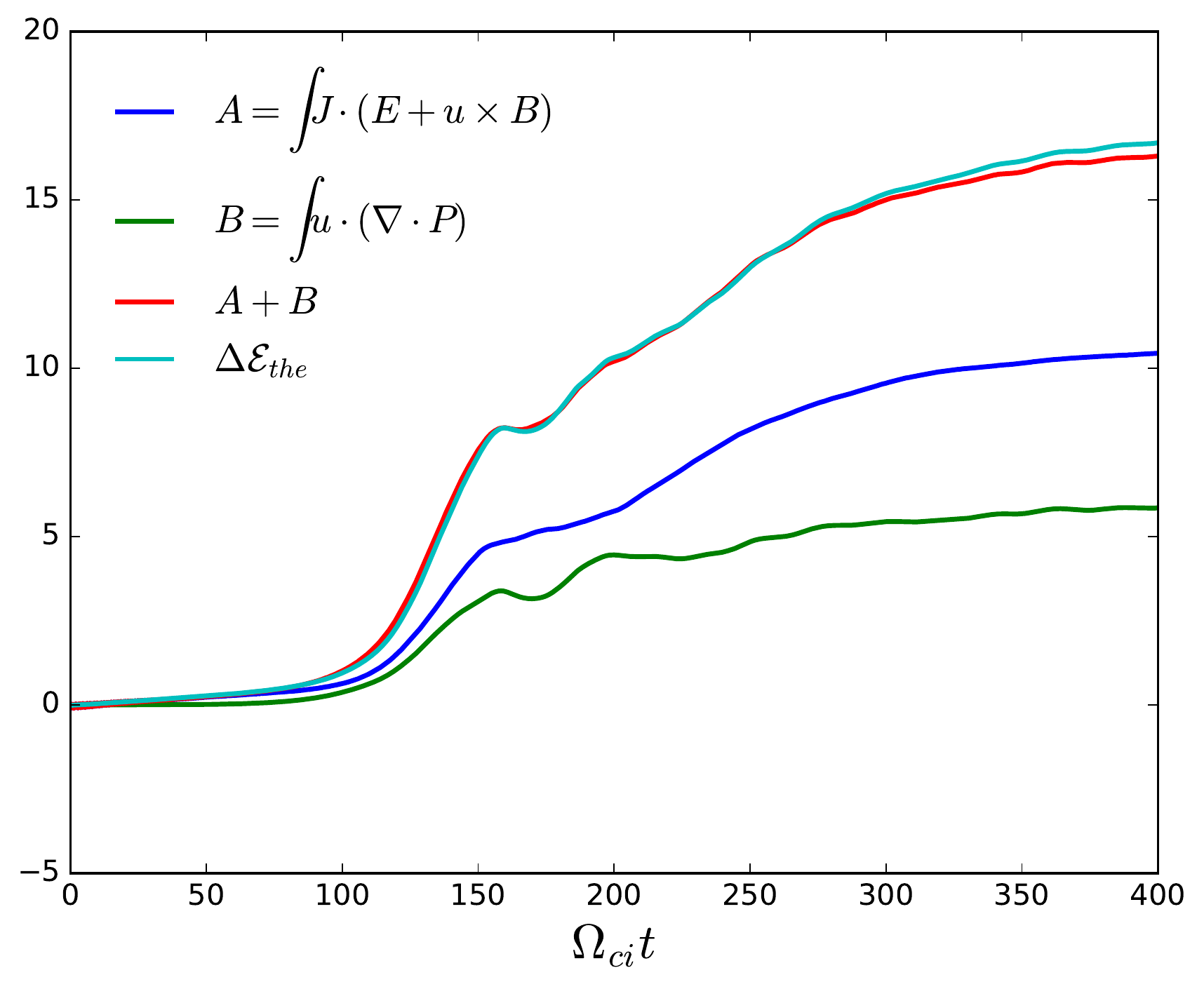}{0.5\textwidth}{(b)}}
\gridline{\fig{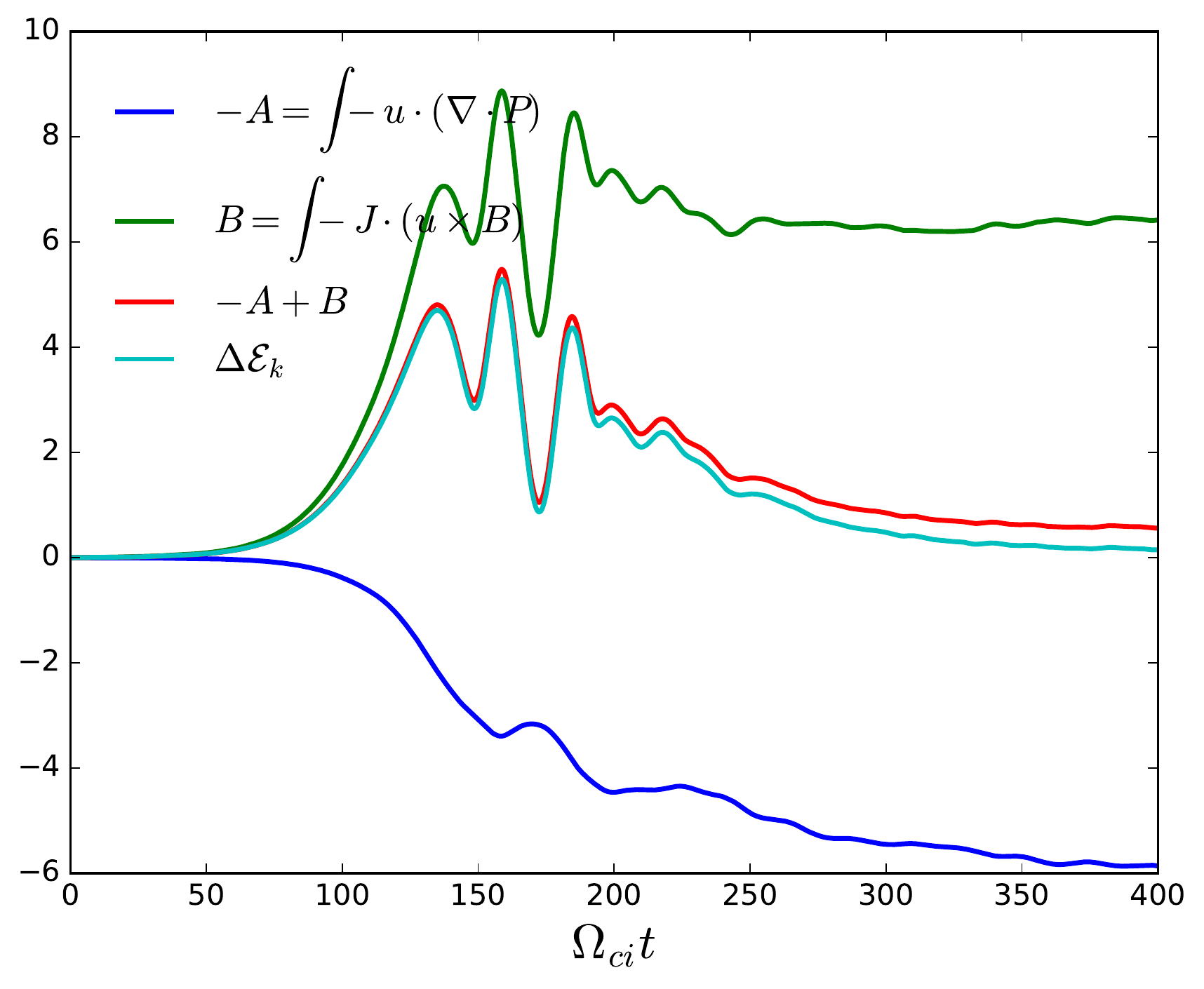}{0.5\textwidth}{(c)}%
          \fig{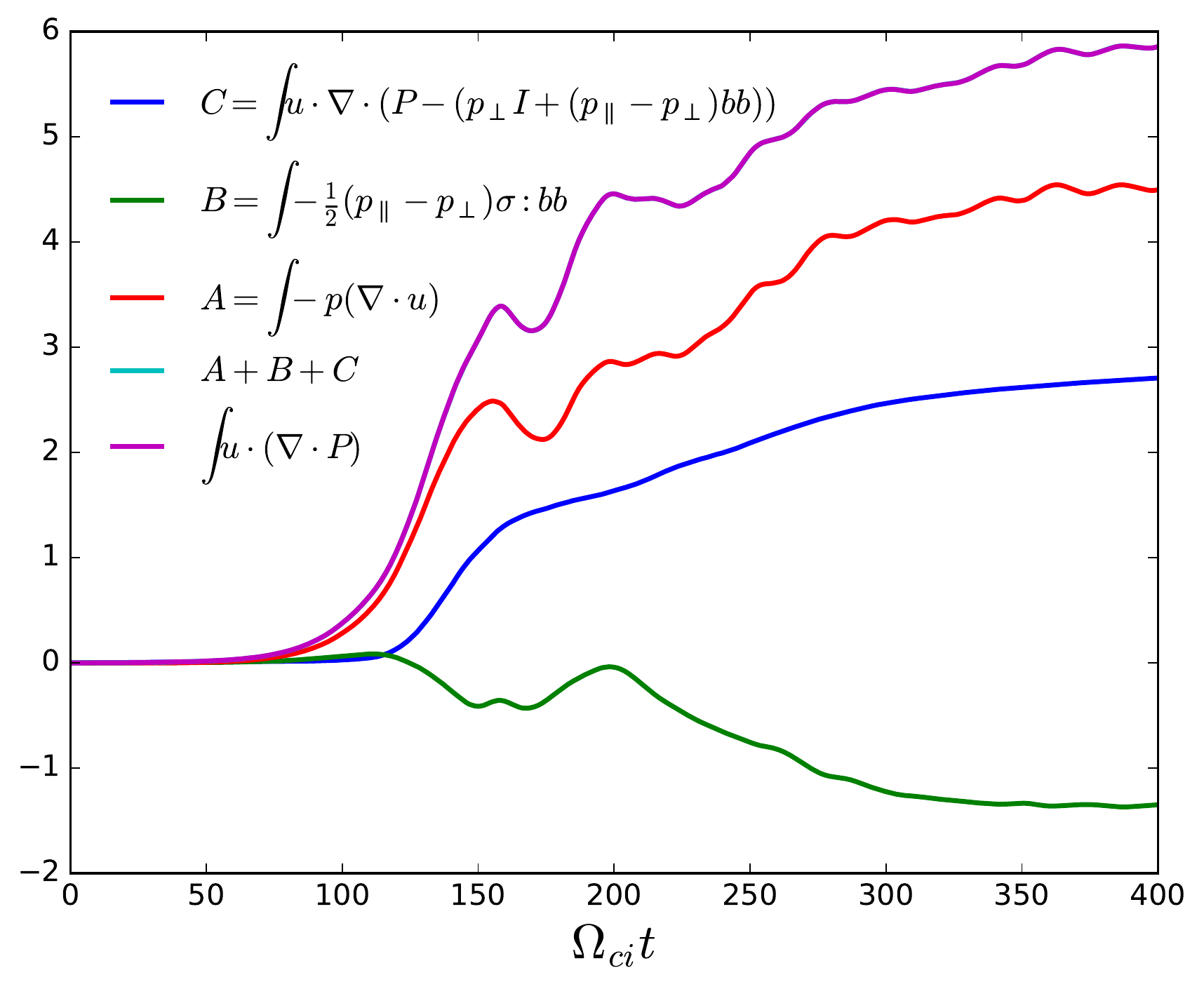}{0.5\textwidth}{(d)}}
\caption{Energy conversion for Run 2. (a) The particle kinetic energy increase $\Delta \mathcal{E}_e$, $\Delta \mathcal{E}_i$, and the work done by the electric field $\iint \boldsymbol{J}_i \cdot \boldsymbol{E} dVdt$, $\iint \boldsymbol{J}_e \cdot \boldsymbol{E} dVdt$. (b) Thermal energy increase, where $A$ and $B$ are the contribution from the non-ideal electric field and the pressure tensor, respectively. (c) Bulk kinetic energy increase, where $A$ and $B$ are the contribution from the pressure tensor and ideal MHD electric field, respectively. (d) The pressure tensor term when separated into the contribution from fluid compression ($A$), flow shear ($B$) and non-gyrotropic pressure ($C$). Note that in (d), the cyan curve overlaps the purple cuve.} \label{fig:energy}
\end{figure}

From a macroscopic point of view, the energy conversion can be understood by taking moments of the collisionless Vlasov equation \citep[e.g.,][]{Zank2014LNP}. In principle, for a collisionless plasma, the kinetic energy gain by a particle species $j$ strictly equals the work done by the electric field. By integrating the energy equation (second moment of the Vlasov equation) and discarding the transport terms,
\begin{equation}\label{eq:etot}
  \frac{d}{dt}\mathcal{E}_{kj} = \int \boldsymbol{J}_j \cdot \boldsymbol{E} dV.
\end{equation}
Here $\mathcal{E}_{kj} = \iint (1/2)m_j v_j^2 f_j d\boldsymbol{v}dV$ is the kinetic energy of species $j$ in a volume, $J_j = n_j q_j \boldsymbol{u}_j$ is the current density of the species, and $\boldsymbol{E}$ is the electric field. We test the energy conversion by integrating the above equation over time in simulations,
\[ \Delta \mathcal{E}_{kj}(t) = \mathcal{E}_{kj}(t) - \mathcal{E}_{kj}(0) = \iint \boldsymbol{J}_j \cdot \boldsymbol{E} dV dt \simeq \sum \boldsymbol{J}_j \cdot \boldsymbol{E} \Delta V \Delta t. \]
The integration is approximated by the summation of the integrand over all numerical cells of volume $\Delta V$ and time steps $\Delta t$. The integration time interval $\Delta t$ needs to be sufficiently small to avoid a large accumulation error. For the first 4 runs, we choose $\Delta t = 0.2\Omega_{ci}^{-1}$, which gives insignificant accumulated errors as shown in Figure \ref{fig:energy} (a). Note that the actual simulation time step $\delta t \simeq 7.8 \times 10^{-2} \omega_{pe}^{-1}$ (or $1.6 \times 10^{-3} \Omega_{ci}^{-1}$ for Run 1--3, and $3.9 \times 10^{-4} \Omega_{ci}^{-1}$ for the Run 4) is much smaller than the $\Delta t$ value. In Run 5, $\Delta t$ is further reduced to $0.05\Omega_{ci}^{-1}$. Although the figure only plots data from Run 2, we note that the other runs also show reasonable agreement between the $\Delta \mathcal{E}$ and $\iint \boldsymbol{J \cdot E}$ curves (the error is slightly larger for Run 3 and 5, but still negligible).

We adopt a single-fluid MHD treatment by combining the moment equations for different species. The total
kinetic energy of particles is separated into a bulk kinetic part and a thermal part, i.e.,
\begin{equation}\label{eq:total_energy}
  \mathcal{E}_k = \frac{1}{2} \rho u^2,\quad \mathcal{E}_{th} = \mathcal{E} - \mathcal{E}_k = \mathcal{E}_i + \mathcal{E}_e - \mathcal{E}_k,
\end{equation}
where $\rho = \rho_i + \rho_e$ is the total mass density, and $\boldsymbol{u} = (\rho_i \boldsymbol{u}_i + \rho_e \boldsymbol{u}_e) / \rho$ is the bulk flow velocity. $\mathcal{E}_i$ and $\mathcal{E}_e$ denote the total kinetic energy of the ions and electrons, i.e., $\mathcal{E}_{i(e)} = \int (1/2)m_{i(e)} v_{i(e)}^2 f_{i(e)} d\boldsymbol{v}_{i(e)}$. Combining the energy equations for ion and electron species yields evolution equations for the bulk kinetic energy and thermal energy \citep[see, e.g.,][]{Birn2012POP},
\begin{equation}\label{eq:ek}
  \frac{\partial}{\partial t}\mathcal{E}_k = \int [\boldsymbol{J \cdot(-u \times B)} - \boldsymbol{u \cdot(\nabla \cdot P)}] dV = \int[\boldsymbol{J} \cdot \boldsymbol{E}_{i} - \boldsymbol{u \cdot (\nabla \cdot P)}] dV;
\end{equation}
\begin{equation}\label{eq:eth}
  \frac{\partial}{\partial t}\mathcal{E}_{the} = \int [\boldsymbol{u \cdot (\nabla \cdot P)} + \boldsymbol{J \cdot(E + u \times B)}] dV = \int [\boldsymbol{u \cdot (\nabla \cdot P)} + \boldsymbol{J} \cdot \boldsymbol{E}_{ni}] dV,
\end{equation}
where the pressure tensor $\boldsymbol{P} = \boldsymbol{P}_i + \boldsymbol{P}_e$ is the sum of the ion and electron pressure tensors that are defined through $\boldsymbol{P}_{i(e)} = \int m_{i(e)}(\boldsymbol{v}_{i(e)}-\boldsymbol{u}_{i(e)})(\boldsymbol{v}_{i(e)}-\boldsymbol{u}_{i(e)}) f_{i(e)}d\boldsymbol{v}_{(i(e)}$. The electric field is decomposed into an ideal MHD part $\boldsymbol{E}_i = -\boldsymbol{u \times B}$ and a non-ideal part $\boldsymbol{E}_{ni} = \boldsymbol{E} - \boldsymbol{E}_{i} = \boldsymbol{E + u \times B}$. Note that the transport terms are already neglected in equations \eqref{eq:ek} and \eqref{eq:eth} since we integrate over the whole volume in an isolated system. A charge separation term $(n_i - n_e)q\boldsymbol{E \cdot u}$ is also neglected since $n_i = n_e$ is usually a good approximation due to the high mobility of electrons. The above equations show that the increase in bulk kinetic energy is due to (a) the ideal electric field $\boldsymbol{E}_i$ and (b) the divergence of the pressure tensor $\boldsymbol{P}$. Similarly, the increase in the thermal energy is due to (a) the non-ideal electric field $\boldsymbol{E}_{ni}$ and also (b) the pressure tensor $\boldsymbol{P}$. The sum of the two equations yields equation \eqref{eq:etot}, showing that the total kinetic energy gain by the particles is due to the work done by electric field. This suggests that the pressure tensor work acts as a bridge that connects the bulk kinetic energy and the thermal energy. A similar analysis is carried out by \citet{Yang2017PoP} for a different problem. They investigate the energy conversion in a 2D turbulent domain for different species, and conclude that the electromagnetic energy is converted to bulk flow energy by the work of electric field, and work by the pressure tensor then channels the bulk flow energy into thermal energy.
Note that Equations \eqref{eq:ek} and \eqref{eq:eth} do not specify the detailed terms of electric field as the generalized Ohm's law.

To illustrate how the pressure tensor term connects with the more commonly used fluid compression, we write the pressure tensor as
\[ \boldsymbol{P} = (\ppar - \pper)\boldsymbol{bb} + \pper\boldsymbol{I} + \boldsymbol{P'} \]
where $\boldsymbol{I}$ is the identity tensor, and $\boldsymbol{b}$ is the unit vector along the local magnetic field. The first two terms make up the diagonal components representing the gyrotropic contribution, and $\boldsymbol{P'}$ consists of the off-diagonal components. The term $\boldsymbol{u}\cdot(\boldsymbol{\nabla \cdot P})$ then becomes
\begin{eqnarray}
  \boldsymbol{u}\cdot(\boldsymbol{\nabla \cdot P}) &=& \boldsymbol{\nabla}\cdot[\pper\boldsymbol{u} + (\ppar - \pper)\boldsymbol{u}\cdot\boldsymbol{bb}] - p\boldsymbol{\nabla \cdot u} - \frac{1}{2}(\ppar - \pper)\boldsymbol{bb : \sigma} \nonumber\\
  && + \boldsymbol{u}\cdot\boldsymbol{\nabla}\cdot\left(\boldsymbol{P} - (\pper\boldsymbol{I} + (\ppar - \pper)\boldsymbol{bb})\right). \label{eq:pressure}
\end{eqnarray}
The integration of the divergence over the whole domain vanishes, so we need only consider the last three terms. The term $p\boldsymbol{\nabla \cdot u}$ clearly represents the normal fluid compression with the scalar pressure $p = (1/3)\mathrm{Tr}(\boldsymbol{P}) = (\ppar + 2\pper)/3$. The term $-(1/2)(\ppar - \pper)\boldsymbol{bb : \sigma}$ can be interpreted as fluid shear, because $\boldsymbol{\sigma}$ is the shear tensor defined as
\[ \sigma_{ij} = \frac{\partial u_i}{\partial x_j} + \frac{\partial u_j}{\partial x_i} - \frac{2}{3}\delta_{ij}\boldsymbol{\nabla \cdot u}. \]
The remaining term comes from the off-diagonal components of the pressure tensor, representing the contribution from the non-gyrotropic pressure. The terms discussed above are plotted for Run 2 in Figures \ref{fig:energy} (b) to (d).

Figures \ref{fig:energy} (b) and (c) show the changes in thermal energy and bulk kinetic energy, respectively. During the earlier phase of the evolution (before $\Omega_{ci}t \sim 150$), as the two islands collide and merge, both the bulk kinetic energy and the thermal energy increase. This can also be seen from panel (a), as both the electron energy and ion energy increase rapidly. Later in the simulations, the bulk kinetic energy starts to decrease, and eventually becomes small compared to the total particle kinetic energy. However, the thermal energy continues increasing while the bulk kinetic energy goes down. At the end of the simulation, almost all particle kinetic energy resides in the plasma thermal energy. The initial increase in the bulk kinetic energy is due to the ideal-MHD electric field, and can be interpreted as the reconnection outflow generated during the coalescence of the two islands. After merging, when magnetic reconnection halts, the outflow is no longer generated. The bulk kinetic energy is then converted to the plasma thermal energy. The conversion of the bulk kinetic energy is through the pressure tensor term $\boldsymbol{u\cdot(\nabla\cdot P)}$, which, as discussed before, can be separated into three parts - fluid compression, shear, and non-gyrotropic pressure. Figure \ref{fig:energy} (d) shows that the fluid compression is the dominant term, while the non-gyrotropic and shear terms combined provide a minor contribution. An interesting feature is that the fluid shear contributes negatively to the pressure tensor work, leading to a decrease of thermal energy. Our preliminary results show that this feature may be related to the ion and electron pressure anisotropy, as suggested by Equation \eqref{eq:pressure}. On the other hand, the non-gyrotropic pressure is mostly supplied by ions, which are easier to demagnetize due to their larger gyroradii. The difference between ion and electron energization is a complicated issue and beyond the scope of this study. Panel (b) also shows that the non-ideal electric field is responsible for most of the thermal energy increase, which is likely due to the relatively strong guide field. This is consistent with previous simulations that indicate that the parallel electric field, which is excluded in the ideal MHD model, tends to dominate the energization when there is a strong guide field \citep{Dahlin2016PoP}.

\subsection{Dependence on the simulation size and mass ratio}

We now discuss the effect of simulation size. In this study, we performed three runs (Run 1, 2 and 3) with box sizes ranging from $\sim 25 d_i$ to $100 d_i$. The evolution time is longer for larger-size simulations, and scales linearly with box size. Despite the difference in size, all the simulation runs show similar features in both general evolution and energy conversion. As shown in Table \ref{tab:energy}, although the amount of energy scales with system size, the percentage changes show only a weak size-dependence. In addition, we compute the energy conversion from various terms as discussed before. The results are listed in Table \ref{tab:conversion}, where the energy conversion is evaluated at the final time of each simulation run. Again, we show that the percentage difference is not significant.
As we discussed before, the numerical heating may affect energy conversion processes to some degree. The test results (1a, 1b and 1c) suggest that the total released magnetic energy decreases during the coalescence process when the level of numerical heating is reduced (Table \ref{tab:energy}). Table \ref{tab:conversion} further suggests that the work of non-ideal electric field $\boldsymbol{E + U\times B}$ decreases more significantly compared to the $-\boldsymbol{U\times B}$ electric field. This may be due to the relatively large noise of the electric field $\boldsymbol{E}$ in the simulations. Nevertheless, the overall behavior of the energy conversion is not affected, and the fact that the total energy increase is very small compared to the energy conversion suggests that the energy exchange between different forms is likely to be physical.

\begin{deluxetable}{cCCCCCCC}[htb!]
\tablecaption{Energy conversion.}
\tablehead{
\colhead{Run \#} & \colhead{$\Delta \mathcal{E}_{i+e}$} & \colhead{$\Delta \mathcal{E}_{k}$} &
\colhead{$\Delta \mathcal{E}_{the}$} & \colhead{$\int \boldsymbol{u\cdot(\nabla\cdot P)}$} &
\colhead{$\int \boldsymbol{J\cdot(E + u\times B)}$} & \colhead{$-\int p\boldsymbol{\nabla \cdot u}$} &
\colhead{$\int \boldsymbol{J\cdot(-u\times B)}$} \\
\colhead{} & \colhead{} & \colhead{} & \colhead{}
& \colhead{(percentage)} & \colhead{(percentage)}
& \colhead{(percentage)} & \colhead{(percentage)}
}
\startdata
1  & +5.38    & +0.06    & +5.38     & 1.88(35\%)   & 3.41(63\%)   & 1.18(22\%)   & 2.00(37\%)   \\
1a & +5.00    & +0.05    & +5.00     & 1.93(39\%)   & 2.96(59\%)   & 1.15(23\%)   & 2.02(40\%)   \\
1b & +4.94    & +0.05    & +4.95     & 1.93(39\%)   & 2.96(59\%)   & 1.16(23\%)   & 1.99(40\%)   \\
1c & +4.80    & +0.04    & +4.81     & 1.98(41\%)   & 2.73(57\%)   & 1.16(24\%)   & 2.05(43\%)   \\
2  & +16.95   & +0.15    & +16.70    & 5.86(35\%)   & 10.45(62\%)  & 4.50(27\%)   & 6.42(38\%)   \\
3  & +61.15   & +0.95    & +59.53    & 21.47(35\%)  & 34.31(56\%)  & 15.33(25\%)  & 25.52(41\%)  \\
4  & +19.83   & +0.18    & +19.57    & 7.80(39\%)   & 11.26(57\%)  & 4.93(25\%)   & 8.27(42\%)   \\
5  & +67.16   & +0.66    & +66.20    & 33.83(50\%)  & 29.95(45\%)  & 19.43(29\%)  & 35.04(52\%)  \\
\enddata
\tablecomments{$\Delta \mathcal{E}_{i+e}$ is the total kinetic energy increase of electrons and ions, $\Delta \mathcal{E}_{k}$ and $\Delta \mathcal{E}_{the}$ are the increase in bulk kinetic energy and thermal energy respectively. The integrals represent contributions from the pressure tensor, ideal-MHD electric field, fluid compression and non-ideal electric field respectively. The unit of energy is $m_e c^2$, and the percentage is relative to $\Delta \mathcal{E}_{i+e}$.} \label{tab:conversion}
\end{deluxetable}

\begin{figure}[!htb]
  \centering
  \includegraphics[width=0.5\linewidth]{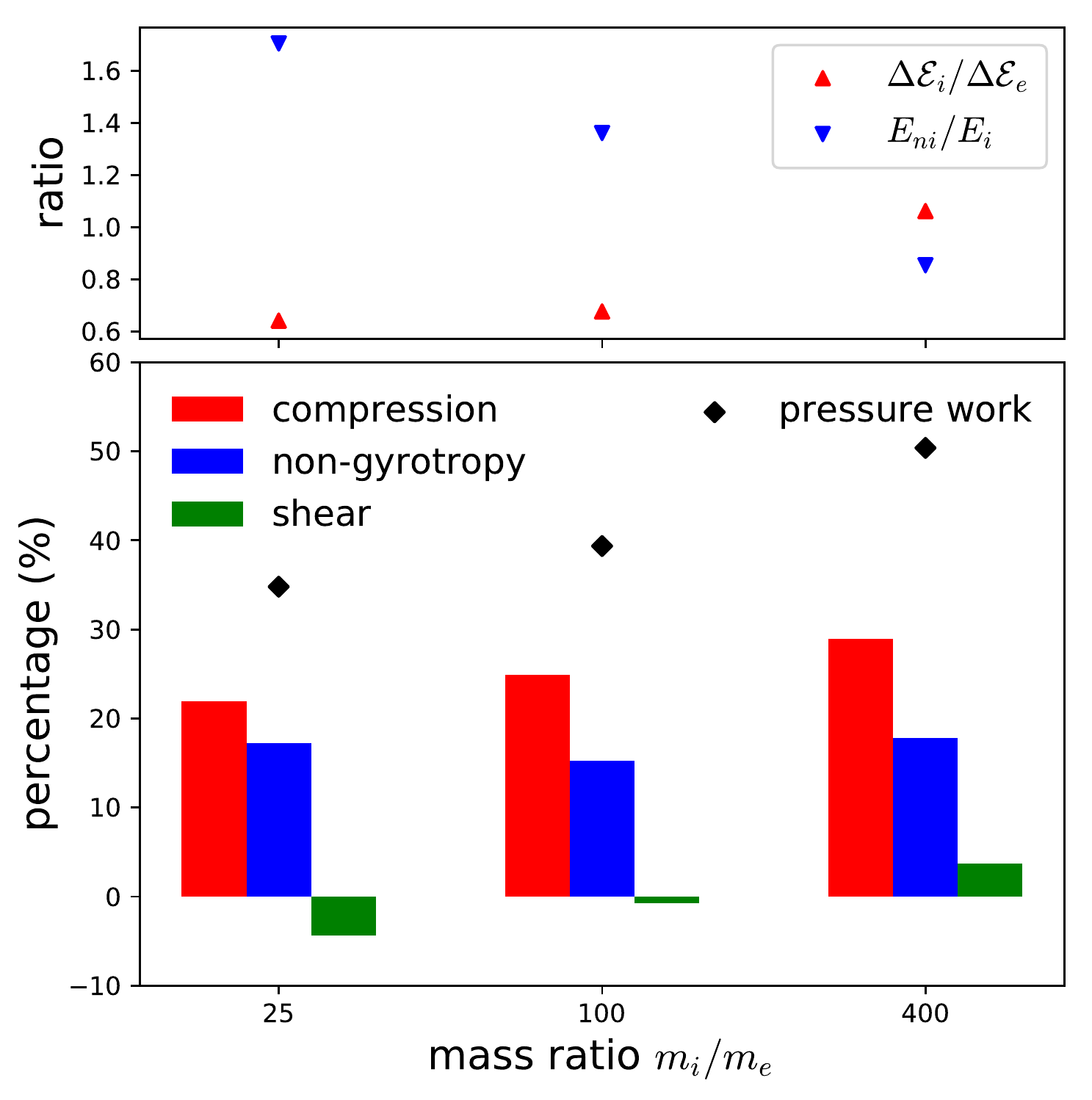}
  \caption{Mass ratio dependence of the energy conversion processes. Data from Runs 1, 4 and 5 are plotted. Top panel: the ratio between ion and electron kinetic energy increase (red triangle); the ratio between the work of non-ideal and ideal electric field (blue reverse triangle). Bottom panel: percentage of the fluid compression (red), shear (green), and non-gyrotropic pressure (blue) contribution relative to the total particle energy gain; the black diamonds are the sum of the three terms, representing the total $\boldsymbol{u \cdot (\nabla\cdot P)}$ contribution.}\label{fig:mass}
\end{figure}

A systematic dependence of the energy conversion on the proton-to-electron mass ratio $m_i / m_e$ is exhibited in Run 1, 4 and 5 with $m_i / m_e = 25, 100$ and 400, respectively. The results are illustrated in Figure \ref{fig:mass}. When the mass ratio is higher, we find that (a) more energy is converted to ion kinetic energy; (b) the contribution from the non-ideal electric field becomes less important and the pressure tensor work contributes more to the increased thermal energy; (c) the fluid compression term contributes more to the energization, and the fluid shear term has also a larger value (more positive). However, the contribution from pressure non-gyrotropy is roughly unchanged with respect to the mass ratio.
Table \ref{tab:energy} shows that the level of numerical heating is larger in the higher mass ratio runs (Run 4 and 5). We caution that this may affect the interpretation of our results. However, as discussed before, an increasing level of numerical heating leads to an increase in the electron energization and the work of non-ideal electric field, which is opposite to what we find in the simulations. This suggests that our results are likely due to physical reasons rather than numerical artifacts. Due to the computational restrictions, only a limited range of simulation parameters are tested. Therefore, the conclusions may only applied to specific systems and are subject to examination against future simulations.

\section{Discussion and Conclusions} \label{sec:diss}

As discussed in \citet{Yang2017PoP}, the work by the pressure tensor transfers large scale flow energy to thermal energy associated with random motions. In our single fluid MHD description, the pressure tensor plays a similar role, although electromagnetic energy is partly converted to thermal energy directly by the non-ideal electric field $\boldsymbol{E}_{ni} = \boldsymbol{E + u \times B}$. By evaluating relative contributions of the pressure tensor work and the fluid compression (shown in table \ref{tab:conversion}), we find that fluid compression indeed dominates over the shear and non-gyrotropic terms for all simulation runs, and is comparable with the contribution from the non-ideal electric field. \citet{Li2018ApJ} discussed the role of fluid compression and shear in electron energization in detail. They also find that fluid compression is the most important contributor in the pressure tensor term. This is consistent with our result, though both the simulation setup and analysis methods are different than our study. Whereas \citet{Li2018ApJ} considered only the perpendicular energization $\boldsymbol{J_{\perp}\cdot E_{\perp}}$ associated with the perpendicular velocity and electric field, we consider the total energization $\boldsymbol{J\cdot E}$ is considered. The parallel energization that is unaccounted for the pressure work is due to the non-ideal electric field. Since compressibility was thought to not to be important and thus neglected in some previous studies regarding magnetic island coalescence \citep[e.g.,][]{Fermo2010PoP, Drake2013ApJL}, the results shown here may be of consequence for developing an energetic particle transport model. We note that in a collisionless system such as our simulations, the particle energization is purely due to the electric field, as illustrated by Equation (1). Therefore, the fluid compression is not an independent energization mechanism, and it happens simultaneously with other mechanisms.

We note that our current analysis focuses on the fluid scale quantities. For an individual particle, it is likely to gain more kinetic energy in a larger scale simulation, because there is more free energy and the particle can experience a longer acceleration time. However, on fluid scales, we do not see differences in plasma energization. Figure \ref{fig:energy} (a) also indicates a difference between ion and electron behaviors, which cannot be studied from a single fluid point of view. We defer a more detailed particle-tracing analysis to future work.

To summarize, we conclude that the plasma energization during magnetic flux rope coalescence is due to the work done by electric field (both ideal and non-ideal), and most of the plasma energy is eventually converted to thermal energy by the pressure tensor. The non-ideal electric field plays an important role in the energy conversion of electromagnetic energy, and fluid compression is the largest contributor to the conversion from bulk flow energy to thermal energy. Finally, we find that our results have only a weak dependence on the simulation size and mass ratio, but an unrealistic mass ratio $m_i / m_e$ may change the details of energy conversion processes.

\acknowledgments
G.P.Z. and S.D. acknowledge the partial support of an NSF DOE grant PHY-1707247 and a NASA grant SV4-84017. This material is based also in part upon work supported by the NSF EPSCoR RII-Track-1 Cooperative Agreement OIA-1655280. F.G. and X.L. acknowledge the support by NASA under grant NNH16AC60I, DOE OFES, and the support by the DOE through the LDRD program at LANL. F.G.ʼs contributions are partly based upon work supported by the U.S. Department of Energy, Office of Fusion Energy Science, under Award Number DE-SC0018240. G.P.Z. acknowledges the generosity of the International Space Science Institute (ISSI) in supporting him through the 2017 Johannes Geiss Fellowship. This work is also partly supported by the International Space Science Institute (ISSI) in the framework of International Team 504 entitled “Current Sheets, Turbulence, Structures and Particle Acceleration in the Heliosphere. S.D. thanks Hui Li and Center for Nonlinear Studies (CNLS) for hosting the visit to Los Alamos National Laboratory (LANL). The simulations are performed at LANL and National Energy Research Scientific Computing Center (NERSC), a DOE Office of Science User Facility supported by the Office of Science of the U.S. Department of Energy under Contract No. DE-AC02-05CH11231.

\bibliographystyle{aasjournal}
\bibliography{Du2018}

\end{document}